\documentclass[twocolumn,showpacs,aps,pra,amsmath,amssymb,superscriptaddress]{revtex4-1}

\usepackage{graphicx}
\usepackage{dcolumn}
\usepackage{bm}
\usepackage{hyperref}
\usepackage{subfigure}

\usepackage{epstopdf}
\usepackage{tikz}

\usepackage{color,bbm}

\newcommand{\beq}{\begin{equation}}
\newcommand{\eeq}{\end{equation}}
\newcommand{\bqa}{\begin{eqnarray}}
\newcommand{\eqa}{\end{eqnarray}}
\newcommand{\nn}{\nonumber}

\newcommand{\rt}[1]{\sqrt{#1}\,}

\newcommand{\bra}[1]{ \langle{#1} |}
\newcommand{\ket}[1]{ |{#1} \rangle}

\newcommand{\sq}[1]{\left[ {#1} \right]}

\newcommand{\tr}[1]{{\rm Tr}\sq{ {#1} }}

\newcommand{\id}{\mathbbm{1}}

\definecolor{maroon}{rgb}{0.7,0,0}

\definecolor{ngreen}{rgb}{0.3,0.7,0.3}

\definecolor{golden}{rgb}{0.8,0.6,0.1}

\begin{document}

\title{Tripartite entanglement measure under local operations and classical communication}

\author{Xiaozhen Ge}
\affiliation{Department of Applied Mathematics, The Hong Kong Polytechnic University, Hong Kong, China}

\author{Lijun Liu}
\email{lljcelia@126.com}
\affiliation{College of Mathematics and Computer Science, Shanxi Normal University, Linfen 041000, China}

\author{Shuming Cheng}
\email{drshuming.cheng@gmail.com}
\affiliation{The Department of Control Science and Engineering, Tongji University, Shanghai 201804, China}
\affiliation{Shanghai Institute of Intelligent Science and Technology, Tongji University, Shanghai 201804, China}
\affiliation{Institute for Advanced Study, Tongji University, Shanghai, 200092, China}

\date{\today}

\begin{abstract}
Multipartite entanglement is an indispensable resource in quantum communication and computation, however, it is a challenging task to faithfully quantify this global property of multipartite quantum systems. In this work, we study the concurrence fill, which admits a geometric interpretation to measure genuine tripartite entanglement for the three-qubit system in [S. Xie {\it et al.}, Phys. Rev. Lett. \textbf{127}. 040403 (2021)]. First, we use the well-known three-tangle and bipartite concurrence to reformulate this quantifier for all pure states. We then construct an explicit example to conclusively show the concurrence fill can be increased under local operations and classical communication (LOCC) {\it on average}, implying it is not an entanglement monotone. Moreover, we give a simple proof of the LOCC-monotonicity of three-tangle and find that the bipartite concurrence and the squared can have distinct performances under the same LOCC. Finally, we propose a reliable monotone to quantify genuine tripartite entanglement, which can also be easily generalised to the multipartite system. Our results shed light on studying genuine entanglement and also reveal the complex structure of multipartite systems.
\end{abstract}


\maketitle
\section{Introduction}
 Entanglement, having no classical counterpart, is of fundamental importance in quantum theory~\cite{Horodecki2009}, and also an essential resource in various quantum information processing tasks, including cryptography~\cite{Ekert1991,Gisin2002}, teleportation~\cite{Bennett1993,Ikram2000}, dense coding~\cite{Bennett1992,Guo2019}, secret sharing~\cite{Hillery1999,Gottesman2000}, metrology~\cite{Vittorio2004,Giovannetti2011}, and computation~\cite{Linden2001}. Thereof, an important problem arises about how to quantitatively measure the degree of entanglement for the quantum system. Typically, one entanglement measure is defined as some non-negative function which maps any quantum state to a real number in the interval $[0,1]$, and satisfies a set of reasonable assumptions~\cite{Vedral1997,Horodecki2009,Guhne2009}, such as being zero for all non-entangled states and invariant under local unitary operations. Within the resource theory of entanglement~\cite{Vidal2000,Chitambar2019}, it is further required to be non-increasing under local operations and classical communication (LOCC) {\it on average}, hence being an entanglement monotone of which entanglement never increases under the free LOCC operations. Correspondingly, numerous entanglement measures and monotones have been developed~\cite{Vedral1998,Horodecki2000,Zyczkowski2001,Plenio2014,Yu2005,Abascal2007}, especially for the low-dimensional bipartite system, such as concurrence~\cite{Hill1997,Wootters1998} and logarithmic negativity~\cite{Plenio2005}.
 
  It is challenging to find proper measures for multipartite entanglement due to the complicated partial separability structure of multipartite systems. Indeed, in order to detect genuine multipartite entanglement which is the key resource in multi-party information tasks, the reliable measure needs to meet extra conditions, like being zero for partial separable states and strictly positive for genuinely entangled states~\cite{Ma2011,Szalay2015}. Thus, there has been some commonly-used quantifiers, including $\alpha$-entanglement entropy~\cite{Szalay2015} and generalized concurrence~\cite{Hiesmayr2008,Hong2012} which can not detect all genuinely entangled states, and few genuine multipartite entanglement monotones~\cite{Emary2004,Sadhukhan2017,Contreras2019,Guo2022,Li2022,Puliyil2022}. In particular, the concurrence fill, with a nice geometric interpretation as the square root of the concurrence triangle area, was recently proposed to measure the degree of genuine entanglement for the three-qubit system~\cite{Xie2021}, together with an experimental test~\cite{Xie2022}. Although it conforms almost all of the necessary conditions mentioned above, a fundamental problem still remains open about whether it is an entanglement monotone, or equivalently, it admits the LOCC-monotonicity. 
 
 \begin{figure*}[!t]
 	\centering
 	
 	\includegraphics[width=
 	0.5\textwidth]{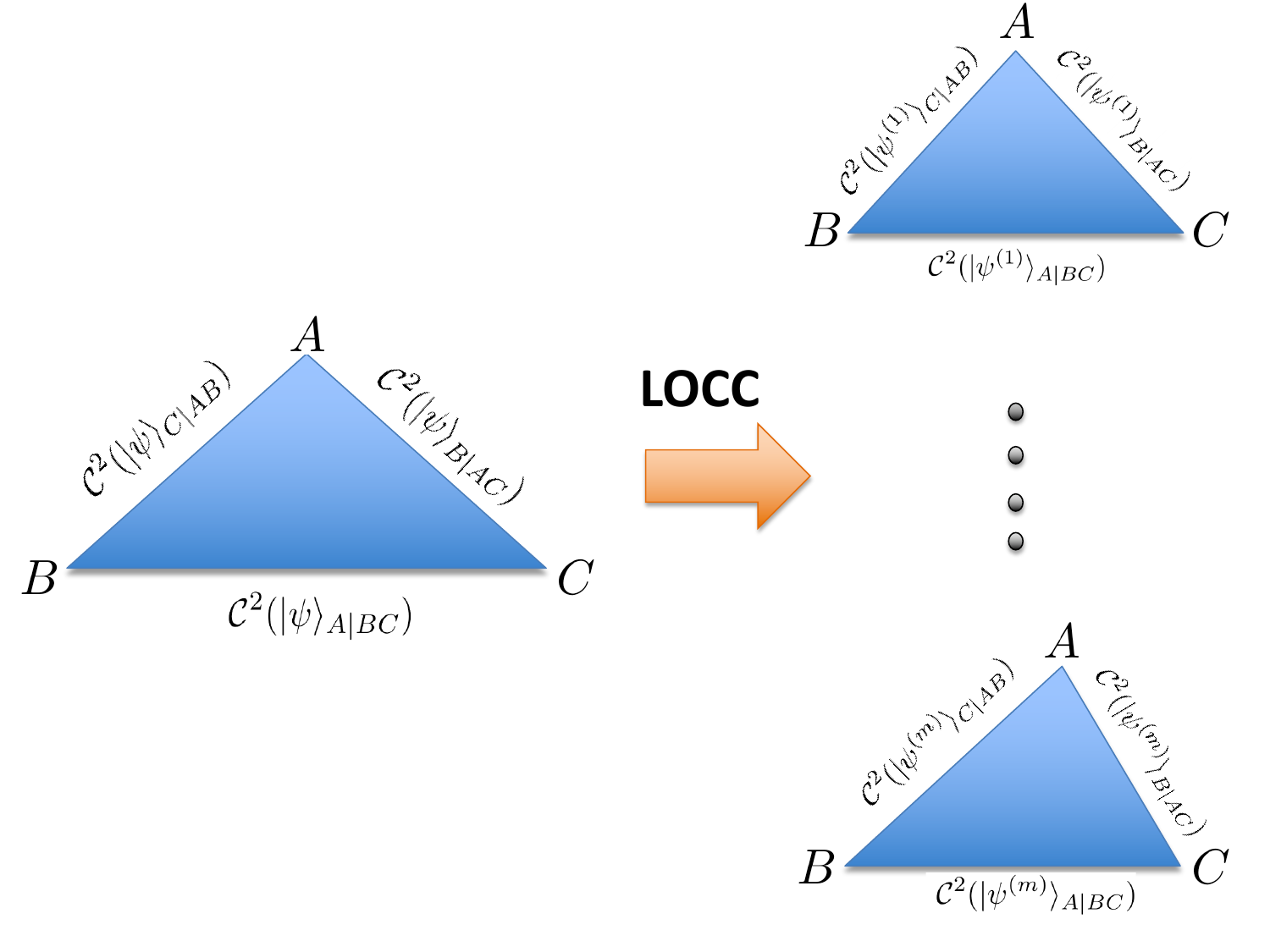}
 	
 	\caption{For any pure three-qubit state $\ket{\psi}_{ABC}$, the concurrence triangle $\Delta ABC$ is composed of three sides described by the three squared one-to-other concurrences $\mathcal{C}^2(\ket{\psi}_{i|jk})$. It leaves open in~\cite{Xie2021} that the square root of the area of this triangle, named as the concurrence fill, is a genuine entanglement monotone, or equivalently, the quantity $\mathcal{F}$ as per (\ref{fill}) is non-increasing under LOCC {\it on average} and thus satisfies the inequality~(\ref{monotonicity}). In this work, we construct an explicit example to show that the concurrence fill can violate the inequality~(\ref{monotonicity}), thus implying it is not an entanglement monotone. }
 	\label{triangle}
 \end{figure*}
 
 Here we first establish the close connections between the concurrence fill and the well-known three-tangle~\cite{Coffman2000} and reduced bipartite concurrences~\cite{Hill1997,Wootters1998,Cohen1998,Divincenzo1999}, of some interest in its own right. Then, we solve the above open problem via an explicit example to show that it can be increased under LOCC {\it on average}, implying it is not an entanglement monotone. Furthermore, a simple proof of the LOCC-monotonicity of three-tangle is presented, which completes the proof in~\cite{Dur2000}. It is also interesting to find that the bipartite concurrence and the squared can have distinct performances under the same LOCC. Finally, we conjecture that the area of the triangle with edges corresponding to the bipartite concurrences is an entanglement monotone and also propose a reliable monotone to quantify genuine tripartite entanglement, which can be easily generalized to the general multipartite systems.

 The rest of this work is structured as follows. Sec.~\ref{sec2} introduces basic notations and useful relations about the concurrence fill for all pure three-qubit states. In Sec.~\ref{sec3}, we address the fundamental issue about whether the concurrence fill is an entanglement monotone or not, and also examine the LOCC-monotonicity of three tangle and bipartite concurrence. Sec.~\ref{alternate} proposes a genuine multipartite entanglement monotone for the multipartite quantum system, and the conclusion is made in Sec.~\ref{sec4}. 
 
\section{Concurrence fill}\label{sec2}

For an arbitrary pure three-qubit state $\ket{\psi}_{ABC}$ shared by three parties $A, B,$ and $C$, denote the concurrence between the bipartition $i$ and $jk$ by $\mathcal{C}(\ket{\psi}_{i|jk})=\rt{2[1-\tr{\rho^2_i}]}$, with $\rho_i={\rm Tr}_{jk}[\ket{\psi}_{ijk}\bra{\psi}]$ and $i, j, k=A, B, C$. It follows from the relation $\mathcal{C}^2(\ket{\psi}_{i|jk})\leq \mathcal{C}^2(\ket{\psi}_{j|ik})+\mathcal{C}^2(\ket{\psi}_{k|ij})$~\cite{Zhu2015} that these three squared one-to-other concurrences can be geometrically interpreted as the lengths of three sides of a triangle, which is called the concurrence triangle $\Delta ABC$ as shown in Fig.~\ref{triangle}. The concurrence fill is defined as the square-root of the area of this concurrence triangle~\cite{Xie2021}
\begin{align}
\mathcal{F}&(\ket{\psi}_{ABC})=\big[\frac{\mathcal{P}_{ABC}}{3}(\mathcal{P}_{ABC}-2\mathcal{C}^2(\ket{\psi}_{A|BC}))\nn\\
&(\mathcal{P}_{ABC}-2\mathcal{C}^2(\ket{\psi}_{B|AC}))(\mathcal{P}_{ABC}-2\mathcal{C}^2(\ket{\psi}_{C|AB}))\big]^{\frac{1}{4}} \label{fill}
\end{align}
with the perimeter
\begin{align}
	\mathcal{P}_{ABC}=\mathcal{C}^2(\ket{\psi}_{A|BC})+\mathcal{C}^2(\ket{\psi}_{B|AC})+\mathcal{C}^2(\ket{\psi}_{C|AB}). \label{perimeter}
	\end{align}
It has been shown in~\cite{Xie2021} the concurrence fill~(\ref{fill}) is useful to quantify genuine entanglement for three-qubit states as it satisfies almost all of the necessary conditions to be an entanglement measure. It is also noted that the perimeter $\mathcal{P}_{ABC}$~(\ref{perimeter}), known as global entanglement~\cite{Meyer2002,Brennen2003}, is a feasible measure, but is not genuine in the sense that it can be non-zero for certain biseparable states.

Further, denoting reduced states $\rho_{ij}={\rm Tr}_k[\ket{\psi}_{ijk}\bra{\psi}]$ and using the relations~\cite{Coffman2000,Cheng2017}
\begin{align}
	\mathcal{C}^2(\ket{\psi}_{i|jk})=2\left(1-\tr{\rho_i^2}\right)=\tau+\mathcal{C}^2(\rho_{ij})+\mathcal{C}^2(\rho_{ik}), \label{CKW}
\end{align}
we are able to derive
\begin{align}
	\mathcal{P}_{ABC}=3\tau+2\left[\mathcal{C}^2(\rho_{AB})+\mathcal{C}^2(\rho_{AC})+\mathcal{C}^2(\rho_{BC})\right], \label{first}
		\end{align}
	and hence
	\begin{align}
		\mathcal{P}_{ABC}-2	\mathcal{C}^2(\ket{\psi}_{i|jk})=\tau+2\mathcal{C}^2(\rho_{jk}). \label{latter}
		\end{align}
	Here $\tau$ is the three-tangle which quantifies the tripartite entanglement~\cite{Coffman2000}, and the concurrence of the reduced two-qubit state $\rho_{ij}$ is given by $\max\{0, \lambda_1-\lambda_2-\lambda_3-\lambda_4\}$ with $\{\lambda_i\}$ the singular values of $X^\top(\sigma_y\otimes\sigma_y)X$ with $\rho_{ij}=XX^\dag$~\cite{Versyraete2001}.
	
	Substituting Eqs.~(\ref{first})-(\ref{latter}) into~(\ref{fill}) immediately leads to
\begin{align}
	&\mathcal{F}(\ket{\psi}_{ABC})\nn\\
	=&\big\{\left[\tau+2\left(\mathcal{C}^2(\rho_{AB})+\mathcal{C}^2(\rho_{AC})+\mathcal{C}^2(\rho_{BC})\right)/3\right]\nn\\
	&(\tau+2\mathcal{C}^2(\rho_{AB}))(\tau+2\mathcal{C}^2(\rho_{AC}))(\tau+2\mathcal{C}^2(\rho_{BC}))\big\}^{\frac{1}{4}}.  \label{area}
\end{align}
It indicates that the concurrence fill can be fully determined by the well-known three tangle and bipartite concurrence. By introducing the concurrence of assistance for a two-qubit state as $\mathcal{C}_a(\rho)=\tr{|X^\top(\sigma_y\otimes\sigma_y)X|}= \lambda_1+\lambda_2+\lambda_3+\lambda_4$~\cite{Laustsen2003} and using $\tau=\mathcal{C}^2_a(\rho_{ij})-\mathcal{C}^2(\rho_{ij})$~\cite{Yu2007}, we can further obtain
\begin{align}
	\mathcal{F}&(\ket{\psi}_{ABC})=\big\{\frac{\mathcal{P}_{ABC}}{3}\left[\mathcal{C}^2(\rho_{AB})+\mathcal{C}^2_a(\rho_{AB})\right]\nn\\
	&\left[\mathcal{C}^2(\rho_{AC})+\mathcal{C}^2_a(\rho_{AC})\right]\left[\mathcal{C}^2(\rho_{BC})+\mathcal{C}^2_a(\rho_{BC})\right]\big\}^{\frac{1}{4}},  \label{concurrence fill}
	\end{align}
with 
\begin{align}
\mathcal{P}_{ABC}= &\,\mathcal{C}^2(\rho_{AB})+\mathcal{C}^2(\rho_{AC})+ \mathcal{C}^2(\rho_{BC})\nn \\
+&\,\mathcal{C}^2_a(\rho_{AB})+\mathcal{C}^2_a(\rho_{AC})+\mathcal{C}^2_a(\rho_{BC}). \nn
\end{align}
This means that the tripartite entanglement quantifier can also be expressed via the reduced bipartite concurrences. Since both two bipartite concurrences and three tangle are entanglement monotones, it is more likely that the concurrence fill shares many similarities with the known entanglement monotones. Thus, Eqs.~(\ref{area}) and~(\ref{concurrence fill}) provide a useful way to investigating the property of $\mathcal{F}$, which will be discussed in the following sections. 

\section{Concurrence fill under LOCCs} \label{sec3}

Within the resource theory of entanglement, the LOCC is the most important class of free operations which cannot generate entanglement. Correspondingly, an important issue about the entanglement monotone is whether it is non-increasing under LOCC {\it on average}. For the above concurrence fill $\mathcal{F}$, as LOCCs do not increase the three lengths of the concurrence triangle $\Delta ABC$, it is natural to draw the conclusion that the corresponding triangle area is non-increasing and hence $\mathcal{F}$ is an entanglement monotone, as conjectured in~\cite{Xie2021}. However, in this section we provide the conclusive evidence to show that this is not the case.

Following from the arguments used to prove the LOCC-monotonicity of three tangle~\cite{Dur2000}, the whole class of LOCCs can be restricted to the binary-outcome positive-operator-valued measures (POVMs) $\{X_1, X_2\}$ acting on the local party $A$, with $X_1^\dag X_1+X^\dag_2X_2=\id$. As a consequence, the problem about whether the concurrence fill is an entanglement monotone reduces to check if it is non-increasing under the above POVMs {\it on average}, or equivalently, the following inequality
\beq
\mathcal{F}(\ket{\psi}_{ABC}) -\left[p_1\,\mathcal{F}(\ket{\psi^{(1)}}_{ABC})+p_2\,\mathcal{F}(\ket{\psi^{(2)}}_{ABC})\right]\geq 0 \label{monotonicity}
\eeq 
holds for all pure states and local measurements, where $\ket{\psi^{(k)}}_{ABC}=X_k\otimes \id\otimes \id \ket{\psi}_{ABC}/\rt{p_k}$ with $p_k=\bra{\psi}_{ABC} X^\dag_kX_k\otimes \id\otimes \id\ket{\psi}_{ABC}$ for $k=1, 2$. If the above inequality is satisfied, then the quantifier $\mathcal{F}$ is an entanglement monotone. Otherwise, it is definitely not. 

Noting further that $\mathcal{F}$ is invariant under the local unitary operations, any pure three-qubit state can be restricted to its standard form~\cite{Acin2000}
\beq
\ket{\psi}_{ABC}=l_0\ket{000}+l_1e^{i\phi}\ket{100}+l_2\ket{101}+l_3\ket{110}+l_4\ket{111}~\label{standard}
\eeq
with $\sum_m l^2_m=1$ and $\phi\in[0, \pi]$, and the measurement operators can be parametrised as $X_i=D_iV$, where
\beq
D_1=\begin{pmatrix}
	\sin\varphi_1 & 0 \\0 & \sin\varphi_2
\end{pmatrix}, 
~D_2=\begin{pmatrix}
\cos\varphi_1 & 0 \\0 & \cos\varphi_2
\end{pmatrix},  \label{diagonal}
\eeq
and 
\beq
V=\begin{pmatrix}
	\cos\psi_1 & -e^{i\psi_2}\sin\psi_1\\
	\sin\psi_1& e^{i\psi_2}\cos\psi_1 
\end{pmatrix} \label{right}
\eeq
with $\varphi_i, \psi_i \in [-\pi, \pi]$. 
Then, the numerical search leads us to find that if the state~(\ref{standard}) is chosen as $l_0|000\rangle+0.096|100\rangle+0.238|101\rangle+0.173|110\rangle$ and the measurements $X_i$ given in Eqs.~(\ref{diagonal}) and~(\ref{right}) are set with $\varphi_1=2\pi/5$, $\varphi_2=\pi/5$, $\psi_1=-\pi/2$, and $\psi_2=-\pi/10$, then the inequality (\ref{monotonicity}) is violated up to
\beq
\mathcal{F}(\ket{\psi}_{ABC})-\sum_{k=1,2}p_k\,\mathcal{F}(\ket{\psi^{(k)}}_{ABC})\approx -0.0086. \label{violation}
\eeq
This immediately implies that the concurrence fill can be increased by some certain LOCC and hence is not a tripartite entanglement monotone.

We remark that two interesting results related to the concurrence fill can also be obtained. The first is a simple proof of the LOCC-monotonicity of three tangle $\tau$ in Eq.~(\ref{area}). Indeed, we are able to prove
 \begin{align}
 &	\tau(\ket{\psi}_{ABC})-\left[p_1\tau(\ket{\psi^{(1)}}_{ABC})+p_2\tau(\ket{\psi^{(2)}}_{ABC})\right] \nn \\
 =&	\frac{\tau(\ket{\psi}_{ABC})}{p_1p_2}\left[p_1p_2-p_2|\det(X_1)|^2-p_1|\det(X_2)|^2\right] \nn \\
   =&\frac{\tau(\ket{\psi}_{ABC})}{p_1p_2}\left[p_1p_2-p_2|\det(D_1)|^2-p_1|\det(D_2)|^2\right] \nn \\
  =&\frac{\tau(\ket{\psi}_{ABC})}{p_1(1-p_1)}(p_1-\sin^2\varphi_1)(\sin^2\varphi_2-p_1) \nn \\
  \geq& 0. \label{tau}
 	\end{align}
The first equality derives from the relation $\tau(\ket{\psi^{(k)}}_{ABC})=\tau(\ket{\psi}_{ABC})|\det(X_k)|^2/p^2_k$~\cite{Dur2000}, the second from $|\det(X_k)|=|\det(D_k)|$, the third from $\sin^2\varphi_k+\cos^2\varphi_k=1$ and $p_1+p_2=1$, and the fourth inequality follows from $\min\{\sin^2\varphi_1, \sin^2\varphi_2\}\leq p_1=\tr{X^\dag_1 X_1\rho_A}\leq \max \{\sin^2\varphi_1, \sin^2\varphi_2\}$ for an arbitrary local state $\rho_A$. This completes the proof that $\tau$ is an entanglement monotone because the LOCC-monotonicity of $\tau^{1/2}$ is proven in~\cite{Dur2000}.

It is also interesting to find that the bipartite concurrence $\mathcal{C}(\rho_{BC})$ and the squared $\mathcal{C}^2(\rho_{BC})$ in the concurrence fill~(\ref{area}) can have distinct performances under the same LOCCs. For example, when each local measurement operator $X_k=\begin{pmatrix}
	x_k & 0 \\0 & y_k
\end{pmatrix}$ acts on the pure three-qubit state $\ket{\psi}_{ABC}$~(\ref{standard}), there are $l^{(k)}_0=x_kl_0/\rt{p_k}$ and $l^{(k)}_m=y_kl_m/\rt{p_k}$ for each $\ket{\psi^{(k)}}_{ABC}$.  Noting that $\mathcal{C}^2(\rho_{BC})=4(l_2^2l_3^2+l^2_1l^2_4-2l_1l_2l_3l_4\cos\phi)$~\cite{Cheng2017} and $p_1=1-p_2=x_1^2l_0^2+y_1^2(1-l_0^2)$, we then obtain
\begin{align}
	&\mathcal{C}^2(\rho_{BC})-\left[p_1\, \mathcal{C}^2(\rho^{(1)}_{BC})+p_2\, \mathcal{C}^2(\rho^{(2)}_{BC})\right]\nn\\
	=&-(x_1^2-y^2_1)^2l^4_0\mathcal{C}^2(\rho_{BC})/p_1p_2\leq 0, \label{squared}
\end{align}
while 
\begin{align}
	&\mathcal{C}(\rho_{BC})-\left[p_1\, \mathcal{C}(\rho^{(1)}_{BC})+p_2\, \mathcal{C}(\rho^{(2)}_{BC})\right]\nn\\
	=&(1-y^2_1-y^2_2)\mathcal{C}(\rho_{BC})=0.
\end{align}
It is evident that if $x_1\neq y_1$ and $l_0, \mathcal{C}(\rho_{BC})\neq 0$, then the squared concurrence is always strictly increased by the above POVMs while the concurrence remains unchanged. We clarify that Eq.~(\ref{squared}) does not contradict with the fact that the squared concurrence is an entanglement monotone for two-qubit states because entanglement can be increased or even generated via local measurements performed by a third party, if a multipartite entangled state is shared.

\section{A reliable multipartite entanglement monotone} \label{alternate}

We have shown via Eq.~(\ref{violation}) that the concurrence fill is not an entanglement monotone, and the main reason lies in the fact that $\mathcal{G}(\ket{\psi}_{ABC})\equiv\tau(\ket{\psi}_{ABC})+2\mathcal{C}^2(\rho_{BC})$ in the concurrence fill~(\ref{area}) can be increased by certain LOCC. This is based on the observation that if two nonnegative quantifiers $\mathcal{E}_j, j=1,2$ are monotonic under LOCCs, i.e., $\mathcal{E}_j(\rho)\geq \sum_kp_k\mathcal{E}(\rho^{(k)})$, then the square-root of their product also obeys the monotonicity as
\begin{align}
	\rt{\mathcal{E}_1(\rho)\mathcal{E}_2(\rho)}&\geq \rt{(\sum_kp_k\mathcal{E}_1(\rho^{(k)}))(\sum_kp_k\mathcal{E}_2(\rho^{(k)}))} \nn \\
	&\geq \sum_k p_k \rt{\mathcal{E}_1(\rho^{(k)})\mathcal{E}_2(\rho^{(k)})}, \label{product}
\end{align}
where the second inequality follows directly from the Cauchy-Schwartz inequality. Hence, if $\mathcal{G}(\ket{\psi}_{ABC})$ admits the LOCC-monotonicity, then using the relation~(\ref{product}) proper times on four terms in the concurrence fill~(\ref{area}) leads to the monotonicity of $\mathcal{F}$. Specifically, one counterexample is given by $\mathcal{G}(\ket{\psi}_{ABC})-p_1\mathcal{G}(\ket{\psi^{(1)}}_{ABC})-p_2\mathcal{G}(\ket{\psi^{(2)}}_{ABC})\approx -0.009$ with $|\psi_{ABC}\rangle=l_0|000\rangle+0.095|100\rangle+0.238|101\rangle+0.086|110\rangle+0.142|111\rangle$ and $\varphi_1=\pi/10$, $\varphi_2=2\pi/5$, $\psi_1=-3\pi/5$ and $\psi_2=-\pi/2$ for the local POVM.

However, the numerical test strongly supports that if the triangle is formed by the bipartite concurrence, then the area of this new triangle 
\begin{align}
	\mathcal{F}^\prime&(\ket{\psi}_{ABC})=\big[\frac{\mathcal{P}^\prime_{ABC}}{3}(\mathcal{P}^\prime_{ABC}-2\mathcal{C}(\ket{\psi}_{A|BC}))\nn\\
	&(\mathcal{P}^\prime_{ABC}-2\mathcal{C}(\ket{\psi}_{B|AC}))(\mathcal{P}^\prime_{ABC}-2\mathcal{C}(\ket{\psi}_{C|AB}))\big]^{\frac{1}{2}} \label{fill2}
\end{align}
with $\mathcal{P}^\prime_{ABC}=\mathcal{C}(\ket{\psi}_{A|BC})+\mathcal{C}(\ket{\psi}_{B|AC})+\mathcal{C}(\ket{\psi}_{C|AB})$, is always non-increasing under LOCC. $\mathcal{F}^\prime$ has a much more natural interpretation than the original concurrence fill~(\ref{fill}), and we conjecture that it is a monotone for genuine entanglement.

We also propose using the square root of the product of three sides to measure tripartite entanglement 
\beq\label{new}
\mathcal{S}(\ket{\psi}_{ABC})=\rt{\mathcal{C}(\ket{\psi}_{A|BC})\mathcal{C}(\ket{\psi}_{B|AC})\mathcal{C}(\ket{\psi}_{C|AB})},
\eeq
for the pure case. It is first easy to show that the above $\mathcal{S}$ is invariant under local unitary operations and permutations of the parties, and is zero if and only if the state is biseparable, i.e., at least one bipartite concurrence is zero. Since the bipartite concurrence and squared ones are monotonic under LOCCs, using inequality~(\ref{product}) twice gives rise to a proof of the LOCC-monotonicity of $\mathcal{S}$. We can also obtain that it achieves the maximal value $1$ with the the Greenberger-Horne-Zeilinger state $\left(\ket{000}+\ket{111}\right)/\rt{2}$, larger than that of the W state $\left(\ket{100}+\ket{010}+\ket{001}\right)/\rt{3}$. Consequently, the quantifier $\mathcal{S}$ as per~(\ref{new}) is a reliable monotone to measure genuine tripartite entanglement, and is readily applied to the mixed states via the convex-roof extension 
$ \mathcal{S}(\rho) =\min_{p_j, \ket{\psi_j}} \sum_j p_j\mathcal{S}(\ket{\psi_j}),$ where the minimisation runs over all ensemble realisations $\{p_j, \ket{\psi_j}\}$ for $\rho$.

It is noted that the genuine entanglement monotone $\mathcal{S}$~(\ref{new}) improves the one based on the geometric mean of bipartite concurrences introduced in~\cite{Li2022} from the order $1/3$ to $1/2$. Moreover, we point out that it can be easily generalised to the multipartite system. For example, for any pure $n$-qubit state denoted by $S$, we can define 
\beq\label{nqubit}
\mathcal{S}(\ket{\psi}_S)=\bigg[\prod_{S_1|S_2}\mathcal{C}(\ket{\psi}_{S_1|S_2})\bigg]^{\frac{1}{2^{n-2}}},
\eeq 
where $S_1|S_2$ denotes any possible bipartition of the $n$-qubit $S$. Using the inequality (\ref{product}) $2^{n-1}-2$ times yields that it is a multipartite entanglement monotone.

\section{Conclusion and discussion }\label{sec4} 

We have studied the concurrence fill~(\ref{fill}) which is introduced to measure genuine entanglement for the three-qubit system. We first build up the close connections between the concurrence fill and the well-known entanglement monotones, including three tangle and bipartite concurrences, as obtained in Eqs.~(\ref{area}) and~(\ref{concurrence fill}). Then, we construct an explicit example~(\ref{violation}) to conclusively show the concurrence fill is not an entanglement monotone, thus answering the fundamental open question left in~\cite{Xie2021}. Moreover, we present a simple and complete proof for the LOCC-monotonicity of three tangle, and also find that the bipartite concurrence can behave different from the squared concurrences under the same LOCC. Finally, we propose a new entanglement monotone~(\ref{product}) for genuine tripartite entanglement, which is readily applied to the multipartite system.

There are many interesting questions left for future work. For example, is it possible to prove that the concurrence fill satisfies $\mathcal{F}(\ket{\psi}_{ABC})\geq \mathcal{F}(\sum p_i\ket{\psi^{(i)}}_{ABC}\bra{\psi^{(i)}})$, thus admitting a less stringent monotonicity under LOCCs? If so, then it could be still an entanglement measure. Otherwise, can we find the counterexample? Instead of the squared concurrence, is it possible to use other bipartite entanglement measures to form a similar triangle and hence to  construct genuine multipartite entanglement measures or monotones which also has a nice geometric interpretation? It is also interesting to investigate the relevant problems about the genuine multipartite nonlocality~\cite{Tavakoli2022} and steering~\cite{Xiang2022}.

\begin{acknowledgments} 
We thank Prof. Shaoming Fei and Yu Guo for useful discussions, and especially acknowledge Songbo Xie for helpful clarifications and discussions about this topic.  S. C. is supported by the Shanghai Municipal Science and Technology Fundamental Project (No. 21JC1405400), the Fundamental Research Funds for the Central Universities (No. 22120220057), and the National Natural Science Foundation of China (No. 12205219, 62088101). L. L. is supported by the National Natural Science Foundation of China (No. 61703254). 
\end{acknowledgments}

\bibliography{Monotonicity}

\begin{thebibliography}{51}%
\makeatletter
\providecommand \@ifxundefined [1]{%
 \@ifx{#1\undefined}
}%
\providecommand \@ifnum [1]{%
 \ifnum #1\expandafter \@firstoftwo
 \else \expandafter \@secondoftwo
 \fi
}%
\providecommand \@ifx [1]{%
 \ifx #1\expandafter \@firstoftwo
 \else \expandafter \@secondoftwo
 \fi
}%
\providecommand \natexlab [1]{#1}%
\providecommand \enquote  [1]{``#1''}%
\providecommand \bibnamefont  [1]{#1}%
\providecommand \bibfnamefont [1]{#1}%
\providecommand \citenamefont [1]{#1}%
\providecommand \href@noop [0]{\@secondoftwo}%
\providecommand \href [0]{\begingroup \@sanitize@url \@href}%
\providecommand \@href[1]{\@@startlink{#1}\@@href}%
\providecommand \@@href[1]{\endgroup#1\@@endlink}%
\providecommand \@sanitize@url [0]{\catcode `\\12\catcode `\$12\catcode
  `\&12\catcode `\#12\catcode `\^12\catcode `\_12\catcode `\%12\relax}%
\providecommand \@@startlink[1]{}%
\providecommand \@@endlink[0]{}%
\providecommand \url  [0]{\begingroup\@sanitize@url \@url }%
\providecommand \@url [1]{\endgroup\@href {#1}{\urlprefix }}%
\providecommand \urlprefix  [0]{URL }%
\providecommand \Eprint [0]{\href }%
\providecommand \doibase [0]{http://dx.doi.org/}%
\providecommand \selectlanguage [0]{\@gobble}%
\providecommand \bibinfo  [0]{\@secondoftwo}%
\providecommand \bibfield  [0]{\@secondoftwo}%
\providecommand \translation [1]{[#1]}%
\providecommand \BibitemOpen [0]{}%
\providecommand \bibitemStop [0]{}%
\providecommand \bibitemNoStop [0]{.\EOS\space}%
\providecommand \EOS [0]{\spacefactor3000\relax}%
\providecommand \BibitemShut  [1]{\csname bibitem#1\endcsname}%
\let\auto@bib@innerbib\@empty
\bibitem [{\citenamefont {Horodecki}\ \emph {et~al.}(2009)\citenamefont
  {Horodecki}, \citenamefont {Horodecki}, \citenamefont {Horodecki},\ and\
  \citenamefont {Horodecki}}]{Horodecki2009}%
  \BibitemOpen
  \bibfield  {author} {\bibinfo {author} {\bibfnamefont {R.}~\bibnamefont
  {Horodecki}}, \bibinfo {author} {\bibfnamefont {P.}~\bibnamefont
  {Horodecki}}, \bibinfo {author} {\bibfnamefont {M.}~\bibnamefont
  {Horodecki}}, \ and\ \bibinfo {author} {\bibfnamefont {K.}~\bibnamefont
  {Horodecki}},\ }\href {\doibase 10.1103/RevModPhys.81.865} {\bibfield
  {journal} {\bibinfo  {journal} {Rev. Mod. Phys.}\ }\textbf {\bibinfo {volume}
  {81}},\ \bibinfo {pages} {865} (\bibinfo {year} {2009})}\BibitemShut
  {NoStop}%
\bibitem [{\citenamefont {Ekert}(1991)}]{Ekert1991}%
  \BibitemOpen
  \bibfield  {author} {\bibinfo {author} {\bibfnamefont {A.~K.}\ \bibnamefont
  {Ekert}},\ }\href {\doibase 10.1103/PhysRevLett.67.661} {\bibfield  {journal}
  {\bibinfo  {journal} {Phys. Rev. Lett.}\ }\textbf {\bibinfo {volume} {67}},\
  \bibinfo {pages} {661} (\bibinfo {year} {1991})}\BibitemShut {NoStop}%
\bibitem [{\citenamefont {Gisin}\ \emph {et~al.}(2002)\citenamefont {Gisin},
  \citenamefont {Ribordy}, \citenamefont {Tittel},\ and\ \citenamefont
  {Zbinden}}]{Gisin2002}%
  \BibitemOpen
  \bibfield  {author} {\bibinfo {author} {\bibfnamefont {N.}~\bibnamefont
  {Gisin}}, \bibinfo {author} {\bibfnamefont {G.}~\bibnamefont {Ribordy}},
  \bibinfo {author} {\bibfnamefont {W.}~\bibnamefont {Tittel}}, \ and\ \bibinfo
  {author} {\bibfnamefont {H.}~\bibnamefont {Zbinden}},\ }\href {\doibase
  10.1103/RevModPhys.74.145} {\bibfield  {journal} {\bibinfo  {journal} {Rev.
  Mod. Phys.}\ }\textbf {\bibinfo {volume} {74}},\ \bibinfo {pages} {145}
  (\bibinfo {year} {2002})}\BibitemShut {NoStop}%
\bibitem [{\citenamefont {Bennett}\ \emph {et~al.}(1993)\citenamefont
  {Bennett}, \citenamefont {Brassard}, \citenamefont {Cr\'epeau}, \citenamefont
  {Jozsa}, \citenamefont {Peres},\ and\ \citenamefont
  {Wootters}}]{Bennett1993}%
  \BibitemOpen
  \bibfield  {author} {\bibinfo {author} {\bibfnamefont {C.~H.}\ \bibnamefont
  {Bennett}}, \bibinfo {author} {\bibfnamefont {G.}~\bibnamefont {Brassard}},
  \bibinfo {author} {\bibfnamefont {C.}~\bibnamefont {Cr\'epeau}}, \bibinfo
  {author} {\bibfnamefont {R.}~\bibnamefont {Jozsa}}, \bibinfo {author}
  {\bibfnamefont {A.}~\bibnamefont {Peres}}, \ and\ \bibinfo {author}
  {\bibfnamefont {W.~K.}\ \bibnamefont {Wootters}},\ }\href {\doibase
  10.1103/PhysRevLett.70.1895} {\bibfield  {journal} {\bibinfo  {journal}
  {Phys. Rev. Lett.}\ }\textbf {\bibinfo {volume} {70}},\ \bibinfo {pages}
  {1895} (\bibinfo {year} {1993})}\BibitemShut {NoStop}%
\bibitem [{\citenamefont {Ikram}\ \emph {et~al.}(2000)\citenamefont {Ikram},
  \citenamefont {Zhu},\ and\ \citenamefont {Zubairy}}]{Ikram2000}%
  \BibitemOpen
  \bibfield  {author} {\bibinfo {author} {\bibfnamefont {M.}~\bibnamefont
  {Ikram}}, \bibinfo {author} {\bibfnamefont {S.-Y.}\ \bibnamefont {Zhu}}, \
  and\ \bibinfo {author} {\bibfnamefont {M.~S.}\ \bibnamefont {Zubairy}},\
  }\href {\doibase 10.1103/PhysRevA.62.022307} {\bibfield  {journal} {\bibinfo
  {journal} {Phys. Rev. A}\ }\textbf {\bibinfo {volume} {62}},\ \bibinfo
  {pages} {022307} (\bibinfo {year} {2000})}\BibitemShut {NoStop}%
\bibitem [{\citenamefont {Bennett}\ and\ \citenamefont
  {Wiesner}(1992)}]{Bennett1992}%
  \BibitemOpen
  \bibfield  {author} {\bibinfo {author} {\bibfnamefont {C.~H.}\ \bibnamefont
  {Bennett}}\ and\ \bibinfo {author} {\bibfnamefont {S.~J.}\ \bibnamefont
  {Wiesner}},\ }\href {\doibase 10.1103/PhysRevLett.69.2881} {\bibfield
  {journal} {\bibinfo  {journal} {Phys. Rev. Lett.}\ }\textbf {\bibinfo
  {volume} {69}},\ \bibinfo {pages} {2881} (\bibinfo {year}
  {1992})}\BibitemShut {NoStop}%
\bibitem [{\citenamefont {Guo}\ \emph {et~al.}(2019)\citenamefont {Guo},
  \citenamefont {Liu}, \citenamefont {Li},\ and\ \citenamefont
  {Guo}}]{Guo2019}%
  \BibitemOpen
  \bibfield  {author} {\bibinfo {author} {\bibfnamefont {Y.}~\bibnamefont
  {Guo}}, \bibinfo {author} {\bibfnamefont {B.-H.}\ \bibnamefont {Liu}},
  \bibinfo {author} {\bibfnamefont {C.-F.}\ \bibnamefont {Li}}, \ and\ \bibinfo
  {author} {\bibfnamefont {G.-C.}\ \bibnamefont {Guo}},\ }\href {\doibase
  https://doi.org/10.1002/qute.201900011} {\bibfield  {journal} {\bibinfo
  {journal} {Advanced Quantum Technologies}\ }\textbf {\bibinfo {volume} {2}},\
  \bibinfo {pages} {1900011} (\bibinfo {year} {2019})}\BibitemShut {NoStop}%
\bibitem [{\citenamefont {Hillery}\ \emph {et~al.}(1999)\citenamefont
  {Hillery}, \citenamefont {Bu\ifmmode~\check{z}\else \v{z}\fi{}ek},\ and\
  \citenamefont {Berthiaume}}]{Hillery1999}%
  \BibitemOpen
  \bibfield  {author} {\bibinfo {author} {\bibfnamefont {M.}~\bibnamefont
  {Hillery}}, \bibinfo {author} {\bibfnamefont {V.}~\bibnamefont
  {Bu\ifmmode~\check{z}\else \v{z}\fi{}ek}}, \ and\ \bibinfo {author}
  {\bibfnamefont {A.}~\bibnamefont {Berthiaume}},\ }\href {\doibase
  10.1103/PhysRevA.59.1829} {\bibfield  {journal} {\bibinfo  {journal} {Phys.
  Rev. A}\ }\textbf {\bibinfo {volume} {59}},\ \bibinfo {pages} {1829}
  (\bibinfo {year} {1999})}\BibitemShut {NoStop}%
\bibitem [{\citenamefont {Gottesman}(2000)}]{Gottesman2000}%
  \BibitemOpen
  \bibfield  {author} {\bibinfo {author} {\bibfnamefont {D.}~\bibnamefont
  {Gottesman}},\ }\href {\doibase 10.1103/PhysRevA.61.042311} {\bibfield
  {journal} {\bibinfo  {journal} {Phys. Rev. A}\ }\textbf {\bibinfo {volume}
  {61}},\ \bibinfo {pages} {042311} (\bibinfo {year} {2000})}\BibitemShut
  {NoStop}%
\bibitem [{\citenamefont {Giovannetti}\ \emph {et~al.}(2004)\citenamefont
  {Giovannetti}, \citenamefont {Lloyd},\ and\ \citenamefont
  {Maccone}}]{Vittorio2004}%
  \BibitemOpen
  \bibfield  {author} {\bibinfo {author} {\bibfnamefont {V.}~\bibnamefont
  {Giovannetti}}, \bibinfo {author} {\bibfnamefont {S.}~\bibnamefont {Lloyd}},
  \ and\ \bibinfo {author} {\bibfnamefont {L.}~\bibnamefont {Maccone}},\ }\href
  {\doibase 10.1126/science.1104149} {\bibfield  {journal} {\bibinfo  {journal}
  {Science}\ }\textbf {\bibinfo {volume} {306}},\ \bibinfo {pages} {1330}
  (\bibinfo {year} {2004})}\BibitemShut {NoStop}%
\bibitem [{\citenamefont {Giovannetti}\ \emph {et~al.}(2011)\citenamefont
  {Giovannetti}, \citenamefont {Lloyd},\ and\ \citenamefont
  {Maccone}}]{Giovannetti2011}%
  \BibitemOpen
  \bibfield  {author} {\bibinfo {author} {\bibfnamefont {V.}~\bibnamefont
  {Giovannetti}}, \bibinfo {author} {\bibfnamefont {S.}~\bibnamefont {Lloyd}},
  \ and\ \bibinfo {author} {\bibfnamefont {L.}~\bibnamefont {Maccone}},\
  }\href@noop {} {\bibfield  {journal} {\bibinfo  {journal} {Nature Photonics}\
  }\textbf {\bibinfo {volume} {5}},\ \bibinfo {pages} {222} (\bibinfo {year}
  {2011})}\BibitemShut {NoStop}%
\bibitem [{\citenamefont {Linden}\ and\ \citenamefont
  {Popescu}(2001)}]{Linden2001}%
  \BibitemOpen
  \bibfield  {author} {\bibinfo {author} {\bibfnamefont {N.}~\bibnamefont
  {Linden}}\ and\ \bibinfo {author} {\bibfnamefont {S.}~\bibnamefont
  {Popescu}},\ }\href {\doibase 10.1103/PhysRevLett.87.047901} {\bibfield
  {journal} {\bibinfo  {journal} {Phys. Rev. Lett.}\ }\textbf {\bibinfo
  {volume} {87}},\ \bibinfo {pages} {047901} (\bibinfo {year}
  {2001})}\BibitemShut {NoStop}%
\bibitem [{\citenamefont {Vedral}\ \emph {et~al.}(1997)\citenamefont {Vedral},
  \citenamefont {Plenio}, \citenamefont {Rippin},\ and\ \citenamefont
  {Knight}}]{Vedral1997}%
  \BibitemOpen
  \bibfield  {author} {\bibinfo {author} {\bibfnamefont {V.}~\bibnamefont
  {Vedral}}, \bibinfo {author} {\bibfnamefont {M.~B.}\ \bibnamefont {Plenio}},
  \bibinfo {author} {\bibfnamefont {M.~A.}\ \bibnamefont {Rippin}}, \ and\
  \bibinfo {author} {\bibfnamefont {P.~L.}\ \bibnamefont {Knight}},\ }\href
  {\doibase 10.1103/PhysRevLett.78.2275} {\bibfield  {journal} {\bibinfo
  {journal} {Phys. Rev. Lett.}\ }\textbf {\bibinfo {volume} {78}},\ \bibinfo
  {pages} {2275} (\bibinfo {year} {1997})}\BibitemShut {NoStop}%
\bibitem [{\citenamefont {G\"{u}hne}\ and\ \citenamefont
  {T\'{o}th}(2009)}]{Guhne2009}%
  \BibitemOpen
  \bibfield  {author} {\bibinfo {author} {\bibfnamefont {O.}~\bibnamefont
  {G\"{u}hne}}\ and\ \bibinfo {author} {\bibfnamefont {G.}~\bibnamefont
  {T\'{o}th}},\ }\href {\doibase https://doi.org/10.1016/j.physrep.2009.02.004}
  {\bibfield  {journal} {\bibinfo  {journal} {Physics Reports}\ }\textbf
  {\bibinfo {volume} {474}},\ \bibinfo {pages} {1} (\bibinfo {year}
  {2009})}\BibitemShut {NoStop}%
\bibitem [{\citenamefont {Vidal}(2000)}]{Vidal2000}%
  \BibitemOpen
  \bibfield  {author} {\bibinfo {author} {\bibfnamefont {G.}~\bibnamefont
  {Vidal}},\ }\href {\doibase 10.1080/09500340008244048} {\bibfield  {journal}
  {\bibinfo  {journal} {Journal of Modern Optics}\ }\textbf {\bibinfo {volume}
  {47}},\ \bibinfo {pages} {355} (\bibinfo {year} {2000})}\BibitemShut
  {NoStop}%
\bibitem [{\citenamefont {Chitambar}\ and\ \citenamefont
  {Gour}(2019)}]{Chitambar2019}%
  \BibitemOpen
  \bibfield  {author} {\bibinfo {author} {\bibfnamefont {E.}~\bibnamefont
  {Chitambar}}\ and\ \bibinfo {author} {\bibfnamefont {G.}~\bibnamefont
  {Gour}},\ }\href {\doibase 10.1103/RevModPhys.91.025001} {\bibfield
  {journal} {\bibinfo  {journal} {Rev. Mod. Phys.}\ }\textbf {\bibinfo {volume}
  {91}},\ \bibinfo {pages} {025001} (\bibinfo {year} {2019})}\BibitemShut
  {NoStop}%
\bibitem [{\citenamefont {Vedral}\ and\ \citenamefont
  {Plenio}(1998)}]{Vedral1998}%
  \BibitemOpen
  \bibfield  {author} {\bibinfo {author} {\bibfnamefont {V.}~\bibnamefont
  {Vedral}}\ and\ \bibinfo {author} {\bibfnamefont {M.~B.}\ \bibnamefont
  {Plenio}},\ }\href {\doibase 10.1103/PhysRevA.57.1619} {\bibfield  {journal}
  {\bibinfo  {journal} {Phys. Rev. A}\ }\textbf {\bibinfo {volume} {57}},\
  \bibinfo {pages} {1619} (\bibinfo {year} {1998})}\BibitemShut {NoStop}%
\bibitem [{\citenamefont {Horodecki}\ \emph {et~al.}(2000)\citenamefont
  {Horodecki}, \citenamefont {Horodecki},\ and\ \citenamefont
  {Horodecki}}]{Horodecki2000}%
  \BibitemOpen
  \bibfield  {author} {\bibinfo {author} {\bibfnamefont {M.}~\bibnamefont
  {Horodecki}}, \bibinfo {author} {\bibfnamefont {P.}~\bibnamefont
  {Horodecki}}, \ and\ \bibinfo {author} {\bibfnamefont {R.}~\bibnamefont
  {Horodecki}},\ }\href {\doibase 10.1103/PhysRevLett.84.2014} {\bibfield
  {journal} {\bibinfo  {journal} {Phys. Rev. Lett.}\ }\textbf {\bibinfo
  {volume} {84}},\ \bibinfo {pages} {2014} (\bibinfo {year}
  {2000})}\BibitemShut {NoStop}%
\bibitem [{\citenamefont {\.{Z}yczkowski}\ \emph {et~al.}(2001)\citenamefont
  {\.{Z}yczkowski}, \citenamefont {Horodecki}, \citenamefont {Horodecki},\ and\
  \citenamefont {Horodecki}}]{Zyczkowski2001}%
  \BibitemOpen
  \bibfield  {author} {\bibinfo {author} {\bibfnamefont {K.}~\bibnamefont
  {\.{Z}yczkowski}}, \bibinfo {author} {\bibfnamefont {P.}~\bibnamefont
  {Horodecki}}, \bibinfo {author} {\bibfnamefont {M.}~\bibnamefont
  {Horodecki}}, \ and\ \bibinfo {author} {\bibfnamefont {R.}~\bibnamefont
  {Horodecki}},\ }\href {\doibase 10.1103/PhysRevA.65.012101} {\bibfield
  {journal} {\bibinfo  {journal} {Phys. Rev. A}\ }\textbf {\bibinfo {volume}
  {65}},\ \bibinfo {pages} {012101} (\bibinfo {year} {2001})}\BibitemShut
  {NoStop}%
\bibitem [{\citenamefont {Plenio}\ and\ \citenamefont
  {Virmani}(2014)}]{Plenio2014}%
  \BibitemOpen
  \bibfield  {author} {\bibinfo {author} {\bibfnamefont {M.~B.}\ \bibnamefont
  {Plenio}}\ and\ \bibinfo {author} {\bibfnamefont {S.~S.}\ \bibnamefont
  {Virmani}},\ }\enquote {\bibinfo {title} {An introduction to entanglement
  theory},}\ in\ \href {\doibase 10.1007/978-3-319-04063-9_8} {\emph {\bibinfo
  {booktitle} {Quantum Information and Coherence}}},\ \bibinfo {editor} {edited
  by\ \bibinfo {editor} {\bibfnamefont {E.}~\bibnamefont {Andersson}}\ and\
  \bibinfo {editor} {\bibfnamefont {P.}~\bibnamefont {{\"O}hberg}}}\ (\bibinfo
  {publisher} {Springer International Publishing},\ \bibinfo {address} {Cham},\
  \bibinfo {year} {2014})\ pp.\ \bibinfo {pages} {173--209}\BibitemShut
  {NoStop}%
\bibitem [{\citenamefont {Yu}\ and\ \citenamefont {Song}(2005)}]{Yu2005}%
  \BibitemOpen
  \bibfield  {author} {\bibinfo {author} {\bibfnamefont {C.-s.}\ \bibnamefont
  {Yu}}\ and\ \bibinfo {author} {\bibfnamefont {H.-s.}\ \bibnamefont {Song}},\
  }\href {\doibase 10.1103/PhysRevA.71.042331} {\bibfield  {journal} {\bibinfo
  {journal} {Phys. Rev. A}\ }\textbf {\bibinfo {volume} {71}},\ \bibinfo
  {pages} {042331} (\bibinfo {year} {2005})}\BibitemShut {NoStop}%
\bibitem [{\citenamefont {Abascal}\ and\ \citenamefont
  {Bj\"{o}rk}(2007)}]{Abascal2007}%
  \BibitemOpen
  \bibfield  {author} {\bibinfo {author} {\bibfnamefont {I.~S.}\ \bibnamefont
  {Abascal}}\ and\ \bibinfo {author} {\bibfnamefont {G.}~\bibnamefont
  {Bj\"{o}rk}},\ }\href {\doibase 10.1103/PhysRevA.75.062317} {\bibfield
  {journal} {\bibinfo  {journal} {Phys. Rev. A}\ }\textbf {\bibinfo {volume}
  {75}},\ \bibinfo {pages} {062317} (\bibinfo {year} {2007})}\BibitemShut
  {NoStop}%
\bibitem [{\citenamefont {Hill}\ and\ \citenamefont
  {Wootters}(1997)}]{Hill1997}%
  \BibitemOpen
  \bibfield  {author} {\bibinfo {author} {\bibfnamefont {S.~A.}\ \bibnamefont
  {Hill}}\ and\ \bibinfo {author} {\bibfnamefont {W.~K.}\ \bibnamefont
  {Wootters}},\ }\href {\doibase 10.1103/PhysRevLett.78.5022} {\bibfield
  {journal} {\bibinfo  {journal} {Phys. Rev. Lett.}\ }\textbf {\bibinfo
  {volume} {78}},\ \bibinfo {pages} {5022} (\bibinfo {year}
  {1997})}\BibitemShut {NoStop}%
\bibitem [{\citenamefont {Wootters}(1998)}]{Wootters1998}%
  \BibitemOpen
  \bibfield  {author} {\bibinfo {author} {\bibfnamefont {W.~K.}\ \bibnamefont
  {Wootters}},\ }\href {\doibase 10.1103/PhysRevLett.80.2245} {\bibfield
  {journal} {\bibinfo  {journal} {Phys. Rev. Lett.}\ }\textbf {\bibinfo
  {volume} {80}},\ \bibinfo {pages} {2245} (\bibinfo {year}
  {1998})}\BibitemShut {NoStop}%
\bibitem [{\citenamefont {Plenio}(2005)}]{Plenio2005}%
  \BibitemOpen
  \bibfield  {author} {\bibinfo {author} {\bibfnamefont {M.~B.}\ \bibnamefont
  {Plenio}},\ }\href {\doibase 10.1103/PhysRevLett.95.090503} {\bibfield
  {journal} {\bibinfo  {journal} {Phys. Rev. Lett.}\ }\textbf {\bibinfo
  {volume} {95}},\ \bibinfo {pages} {090503} (\bibinfo {year}
  {2005})}\BibitemShut {NoStop}%
\bibitem [{\citenamefont {Ma}\ \emph {et~al.}(2011)\citenamefont {Ma},
  \citenamefont {Chen}, \citenamefont {Chen}, \citenamefont {Spengler},
  \citenamefont {Gabriel},\ and\ \citenamefont {Huber}}]{Ma2011}%
  \BibitemOpen
  \bibfield  {author} {\bibinfo {author} {\bibfnamefont {Z.-H.}\ \bibnamefont
  {Ma}}, \bibinfo {author} {\bibfnamefont {Z.-H.}\ \bibnamefont {Chen}},
  \bibinfo {author} {\bibfnamefont {J.-L.}\ \bibnamefont {Chen}}, \bibinfo
  {author} {\bibfnamefont {C.}~\bibnamefont {Spengler}}, \bibinfo {author}
  {\bibfnamefont {A.}~\bibnamefont {Gabriel}}, \ and\ \bibinfo {author}
  {\bibfnamefont {M.}~\bibnamefont {Huber}},\ }\href {\doibase
  10.1103/PhysRevA.83.062325} {\bibfield  {journal} {\bibinfo  {journal} {Phys.
  Rev. A}\ }\textbf {\bibinfo {volume} {83}},\ \bibinfo {pages} {062325}
  (\bibinfo {year} {2011})}\BibitemShut {NoStop}%
\bibitem [{\citenamefont {Szalay}(2015)}]{Szalay2015}%
  \BibitemOpen
  \bibfield  {author} {\bibinfo {author} {\bibfnamefont {S.}~\bibnamefont
  {Szalay}},\ }\href {\doibase 10.1103/PhysRevA.92.042329} {\bibfield
  {journal} {\bibinfo  {journal} {Phys. Rev. A}\ }\textbf {\bibinfo {volume}
  {92}},\ \bibinfo {pages} {042329} (\bibinfo {year} {2015})}\BibitemShut
  {NoStop}%
\bibitem [{\citenamefont {Hiesmayr}\ and\ \citenamefont
  {Huber}(2008)}]{Hiesmayr2008}%
  \BibitemOpen
  \bibfield  {author} {\bibinfo {author} {\bibfnamefont {B.~C.}\ \bibnamefont
  {Hiesmayr}}\ and\ \bibinfo {author} {\bibfnamefont {M.}~\bibnamefont
  {Huber}},\ }\href {\doibase 10.1103/PhysRevA.78.012342} {\bibfield  {journal}
  {\bibinfo  {journal} {Phys. Rev. A}\ }\textbf {\bibinfo {volume} {78}},\
  \bibinfo {pages} {012342} (\bibinfo {year} {2008})}\BibitemShut {NoStop}%
\bibitem [{\citenamefont {Hong}\ \emph {et~al.}(2012)\citenamefont {Hong},
  \citenamefont {Gao},\ and\ \citenamefont {Yan}}]{Hong2012}%
  \BibitemOpen
  \bibfield  {author} {\bibinfo {author} {\bibfnamefont {Y.}~\bibnamefont
  {Hong}}, \bibinfo {author} {\bibfnamefont {T.}~\bibnamefont {Gao}}, \ and\
  \bibinfo {author} {\bibfnamefont {F.}~\bibnamefont {Yan}},\ }\href {\doibase
  10.1103/PhysRevA.86.062323} {\bibfield  {journal} {\bibinfo  {journal} {Phys.
  Rev. A}\ }\textbf {\bibinfo {volume} {86}},\ \bibinfo {pages} {062323}
  (\bibinfo {year} {2012})}\BibitemShut {NoStop}%
\bibitem [{\citenamefont {Emary}\ and\ \citenamefont
  {Beenakker}(2004)}]{Emary2004}%
  \BibitemOpen
  \bibfield  {author} {\bibinfo {author} {\bibfnamefont {C.}~\bibnamefont
  {Emary}}\ and\ \bibinfo {author} {\bibfnamefont {C.~W.~J.}\ \bibnamefont
  {Beenakker}},\ }\href {\doibase 10.1103/PhysRevA.69.032317} {\bibfield
  {journal} {\bibinfo  {journal} {Phys. Rev. A}\ }\textbf {\bibinfo {volume}
  {69}},\ \bibinfo {pages} {032317} (\bibinfo {year} {2004})}\BibitemShut
  {NoStop}%
\bibitem [{\citenamefont {Sadhukhan}\ \emph {et~al.}(2017)\citenamefont
  {Sadhukhan}, \citenamefont {Roy}, \citenamefont {Pal}, \citenamefont
  {Rakshit}, \citenamefont {Sen(De)},\ and\ \citenamefont
  {Sen}}]{Sadhukhan2017}%
  \BibitemOpen
  \bibfield  {author} {\bibinfo {author} {\bibfnamefont {D.}~\bibnamefont
  {Sadhukhan}}, \bibinfo {author} {\bibfnamefont {S.~S.}\ \bibnamefont {Roy}},
  \bibinfo {author} {\bibfnamefont {A.~K.}\ \bibnamefont {Pal}}, \bibinfo
  {author} {\bibfnamefont {D.}~\bibnamefont {Rakshit}}, \bibinfo {author}
  {\bibfnamefont {A.}~\bibnamefont {Sen(De)}}, \ and\ \bibinfo {author}
  {\bibfnamefont {U.}~\bibnamefont {Sen}},\ }\href {\doibase
  10.1103/PhysRevA.95.022301} {\bibfield  {journal} {\bibinfo  {journal} {Phys.
  Rev. A}\ }\textbf {\bibinfo {volume} {95}},\ \bibinfo {pages} {022301}
  (\bibinfo {year} {2017})}\BibitemShut {NoStop}%
\bibitem [{\citenamefont {Contreras-Tejada}\ \emph {et~al.}(2019)\citenamefont
  {Contreras-Tejada}, \citenamefont {Palazuelos},\ and\ \citenamefont
  {de~Vicente}}]{Contreras2019}%
  \BibitemOpen
  \bibfield  {author} {\bibinfo {author} {\bibfnamefont {P.}~\bibnamefont
  {Contreras-Tejada}}, \bibinfo {author} {\bibfnamefont {C.}~\bibnamefont
  {Palazuelos}}, \ and\ \bibinfo {author} {\bibfnamefont {J.~I.}\ \bibnamefont
  {de~Vicente}},\ }\href {\doibase 10.1103/PhysRevLett.122.120503} {\bibfield
  {journal} {\bibinfo  {journal} {Phys. Rev. Lett.}\ }\textbf {\bibinfo
  {volume} {122}},\ \bibinfo {pages} {120503} (\bibinfo {year}
  {2019})}\BibitemShut {NoStop}%
\bibitem [{\citenamefont {Guo}\ \emph {et~al.}(2022)\citenamefont {Guo},
  \citenamefont {Jia}, \citenamefont {Li},\ and\ \citenamefont
  {Huang}}]{Guo2022}%
  \BibitemOpen
  \bibfield  {author} {\bibinfo {author} {\bibfnamefont {Y.}~\bibnamefont
  {Guo}}, \bibinfo {author} {\bibfnamefont {Y.}~\bibnamefont {Jia}}, \bibinfo
  {author} {\bibfnamefont {X.}~\bibnamefont {Li}}, \ and\ \bibinfo {author}
  {\bibfnamefont {L.}~\bibnamefont {Huang}},\ }\href {\doibase
  10.1088/1751-8121/ac5649} {\bibfield  {journal} {\bibinfo  {journal} {Journal
  of Physics A: Mathematical and Theoretical}\ }\textbf {\bibinfo {volume}
  {55}},\ \bibinfo {pages} {145303} (\bibinfo {year} {2022})}\BibitemShut
  {NoStop}%
\bibitem [{\citenamefont {Li}\ and\ \citenamefont {Shang}(2022)}]{Li2022}%
  \BibitemOpen
  \bibfield  {author} {\bibinfo {author} {\bibfnamefont {Y.}~\bibnamefont
  {Li}}\ and\ \bibinfo {author} {\bibfnamefont {J.}~\bibnamefont {Shang}},\
  }\href {\doibase 10.1103/PhysRevResearch.4.023059} {\bibfield  {journal}
  {\bibinfo  {journal} {Phys. Rev. Research}\ }\textbf {\bibinfo {volume}
  {4}},\ \bibinfo {pages} {023059} (\bibinfo {year} {2022})}\BibitemShut
  {NoStop}%
\bibitem [{\citenamefont {Puliyil}\ \emph {et~al.}(2022)\citenamefont
  {Puliyil}, \citenamefont {Banik},\ and\ \citenamefont
  {Alimuddin}}]{Puliyil2022}%
  \BibitemOpen
  \bibfield  {author} {\bibinfo {author} {\bibfnamefont {S.}~\bibnamefont
  {Puliyil}}, \bibinfo {author} {\bibfnamefont {M.}~\bibnamefont {Banik}}, \
  and\ \bibinfo {author} {\bibfnamefont {M.}~\bibnamefont {Alimuddin}},\ }\href
  {\doibase 10.1103/PhysRevLett.129.070601} {\bibfield  {journal} {\bibinfo
  {journal} {Phys. Rev. Lett.}\ }\textbf {\bibinfo {volume} {129}},\ \bibinfo
  {pages} {070601} (\bibinfo {year} {2022})}\BibitemShut {NoStop}%
\bibitem [{\citenamefont {Xie}\ and\ \citenamefont {Eberly}(2021)}]{Xie2021}%
  \BibitemOpen
  \bibfield  {author} {\bibinfo {author} {\bibfnamefont {S.}~\bibnamefont
  {Xie}}\ and\ \bibinfo {author} {\bibfnamefont {J.~H.}\ \bibnamefont
  {Eberly}},\ }\href {\doibase 10.1103/PhysRevLett.127.040403} {\bibfield
  {journal} {\bibinfo  {journal} {Phys. Rev. Lett.}\ }\textbf {\bibinfo
  {volume} {127}},\ \bibinfo {pages} {040403} (\bibinfo {year}
  {2021})}\BibitemShut {NoStop}%
\bibitem [{\citenamefont {Xie}\ \emph {et~al.}(2022)\citenamefont {Xie},
  \citenamefont {Zhao}, \citenamefont {Zhang}, \citenamefont {Huang},
  \citenamefont {Li}, \citenamefont {Guo},\ and\ \citenamefont
  {Eberly}}]{Xie2022}%
  \BibitemOpen
  \bibfield  {author} {\bibinfo {author} {\bibfnamefont {S.}~\bibnamefont
  {Xie}}, \bibinfo {author} {\bibfnamefont {Y.-Y.}\ \bibnamefont {Zhao}},
  \bibinfo {author} {\bibfnamefont {C.}~\bibnamefont {Zhang}}, \bibinfo
  {author} {\bibfnamefont {Y.-F.}\ \bibnamefont {Huang}}, \bibinfo {author}
  {\bibfnamefont {C.-F.}\ \bibnamefont {Li}}, \bibinfo {author} {\bibfnamefont
  {G.-C.}\ \bibnamefont {Guo}}, \ and\ \bibinfo {author} {\bibfnamefont
  {J.~H.}\ \bibnamefont {Eberly}},\ }\href@noop {} {\bibfield  {journal}
  {\bibinfo  {journal} {arXiv preprint arXiv:2207.07584}\ } (\bibinfo {year}
  {2022})}\BibitemShut {NoStop}%
\bibitem [{\citenamefont {Coffman}\ \emph {et~al.}(2000)\citenamefont
  {Coffman}, \citenamefont {Kundu},\ and\ \citenamefont
  {Wootters}}]{Coffman2000}%
  \BibitemOpen
  \bibfield  {author} {\bibinfo {author} {\bibfnamefont {V.}~\bibnamefont
  {Coffman}}, \bibinfo {author} {\bibfnamefont {J.}~\bibnamefont {Kundu}}, \
  and\ \bibinfo {author} {\bibfnamefont {W.~K.}\ \bibnamefont {Wootters}},\
  }\href {\doibase 10.1103/PhysRevA.61.052306} {\bibfield  {journal} {\bibinfo
  {journal} {Phys. Rev. A}\ }\textbf {\bibinfo {volume} {61}},\ \bibinfo
  {pages} {052306} (\bibinfo {year} {2000})}\BibitemShut {NoStop}%
\bibitem [{\citenamefont {Cohen}(1998)}]{Cohen1998}%
  \BibitemOpen
  \bibfield  {author} {\bibinfo {author} {\bibfnamefont {O.}~\bibnamefont
  {Cohen}},\ }\href {\doibase 10.1103/PhysRevLett.80.2493} {\bibfield
  {journal} {\bibinfo  {journal} {Phys. Rev. Lett.}\ }\textbf {\bibinfo
  {volume} {80}},\ \bibinfo {pages} {2493} (\bibinfo {year}
  {1998})}\BibitemShut {NoStop}%
\bibitem [{\citenamefont {DiVincenzo}\ \emph {et~al.}(1999)\citenamefont
  {DiVincenzo}, \citenamefont {Fuchs}, \citenamefont {Mabuchi}, \citenamefont
  {Smolin}, \citenamefont {Thapliyal},\ and\ \citenamefont
  {Uhlmann}}]{Divincenzo1999}%
  \BibitemOpen
  \bibfield  {author} {\bibinfo {author} {\bibfnamefont {D.~P.}\ \bibnamefont
  {DiVincenzo}}, \bibinfo {author} {\bibfnamefont {C.~A.}\ \bibnamefont
  {Fuchs}}, \bibinfo {author} {\bibfnamefont {H.}~\bibnamefont {Mabuchi}},
  \bibinfo {author} {\bibfnamefont {J.~A.}\ \bibnamefont {Smolin}}, \bibinfo
  {author} {\bibfnamefont {A.}~\bibnamefont {Thapliyal}}, \ and\ \bibinfo
  {author} {\bibfnamefont {A.}~\bibnamefont {Uhlmann}},\ }in\ \href@noop {}
  {\emph {\bibinfo {booktitle} {Quantum Computing and Quantum
  Communications}}},\ \bibinfo {editor} {edited by\ \bibinfo {editor}
  {\bibfnamefont {C.~P.}\ \bibnamefont {Williams}}}\ (\bibinfo  {publisher}
  {Springer Berlin Heidelberg},\ \bibinfo {address} {Berlin, Heidelberg},\
  \bibinfo {year} {1999})\ pp.\ \bibinfo {pages} {247--257}\BibitemShut
  {NoStop}%
\bibitem [{\citenamefont {D\"ur}\ \emph {et~al.}(2000)\citenamefont {D\"ur},
  \citenamefont {Vidal},\ and\ \citenamefont {Cirac}}]{Dur2000}%
  \BibitemOpen
  \bibfield  {author} {\bibinfo {author} {\bibfnamefont {W.}~\bibnamefont
  {D\"ur}}, \bibinfo {author} {\bibfnamefont {G.}~\bibnamefont {Vidal}}, \ and\
  \bibinfo {author} {\bibfnamefont {J.~I.}\ \bibnamefont {Cirac}},\ }\href
  {\doibase 10.1103/PhysRevA.62.062314} {\bibfield  {journal} {\bibinfo
  {journal} {Phys. Rev. A}\ }\textbf {\bibinfo {volume} {62}},\ \bibinfo
  {pages} {062314} (\bibinfo {year} {2000})}\BibitemShut {NoStop}%
\bibitem [{\citenamefont {Zhu}\ and\ \citenamefont {Fei}(2015)}]{Zhu2015}%
  \BibitemOpen
  \bibfield  {author} {\bibinfo {author} {\bibfnamefont {X.-N.}\ \bibnamefont
  {Zhu}}\ and\ \bibinfo {author} {\bibfnamefont {S.-M.}\ \bibnamefont {Fei}},\
  }\href {\doibase 10.1103/PhysRevA.92.062345} {\bibfield  {journal} {\bibinfo
  {journal} {Phys. Rev. A}\ }\textbf {\bibinfo {volume} {92}},\ \bibinfo
  {pages} {062345} (\bibinfo {year} {2015})}\BibitemShut {NoStop}%
\bibitem [{\citenamefont {Meyer}\ and\ \citenamefont
  {Wallach}(2002)}]{Meyer2002}%
  \BibitemOpen
  \bibfield  {author} {\bibinfo {author} {\bibfnamefont {D.~A.}\ \bibnamefont
  {Meyer}}\ and\ \bibinfo {author} {\bibfnamefont {N.~R.}\ \bibnamefont
  {Wallach}},\ }\href {\doibase 10.1063/1.1497700} {\bibfield  {journal}
  {\bibinfo  {journal} {Journal of Mathematical Physics}\ }\textbf {\bibinfo
  {volume} {43}},\ \bibinfo {pages} {4273} (\bibinfo {year}
  {2002})}\BibitemShut {NoStop}%
\bibitem [{\citenamefont {Brennen}(2003)}]{Brennen2003}%
  \BibitemOpen
  \bibfield  {author} {\bibinfo {author} {\bibfnamefont {G.~K.}\ \bibnamefont
  {Brennen}},\ }\href@noop {} {\bibfield  {journal} {\bibinfo  {journal}
  {Quantum Info. Comput.}\ }\textbf {\bibinfo {volume} {3}},\ \bibinfo {pages}
  {619–626} (\bibinfo {year} {2003})}\BibitemShut {NoStop}%
\bibitem [{\citenamefont {Cheng}\ and\ \citenamefont {Hall}(2017)}]{Cheng2017}%
  \BibitemOpen
  \bibfield  {author} {\bibinfo {author} {\bibfnamefont {S.}~\bibnamefont
  {Cheng}}\ and\ \bibinfo {author} {\bibfnamefont {M.~J.~W.}\ \bibnamefont
  {Hall}},\ }\href {\doibase 10.1103/PhysRevLett.118.010401} {\bibfield
  {journal} {\bibinfo  {journal} {Phys. Rev. Lett.}\ }\textbf {\bibinfo
  {volume} {118}},\ \bibinfo {pages} {010401} (\bibinfo {year}
  {2017})}\BibitemShut {NoStop}%
\bibitem [{\citenamefont {Verstraete}\ \emph {et~al.}(2001)\citenamefont
  {Verstraete}, \citenamefont {Audenaert},\ and\ \citenamefont
  {De~Moor}}]{Versyraete2001}%
  \BibitemOpen
  \bibfield  {author} {\bibinfo {author} {\bibfnamefont {F.}~\bibnamefont
  {Verstraete}}, \bibinfo {author} {\bibfnamefont {K.}~\bibnamefont
  {Audenaert}}, \ and\ \bibinfo {author} {\bibfnamefont {B.}~\bibnamefont
  {De~Moor}},\ }\href {\doibase 10.1103/PhysRevA.64.012316} {\bibfield
  {journal} {\bibinfo  {journal} {Phys. Rev. A}\ }\textbf {\bibinfo {volume}
  {64}},\ \bibinfo {pages} {012316} (\bibinfo {year} {2001})}\BibitemShut
  {NoStop}%
\bibitem [{\citenamefont {Laustsen}\ \emph {et~al.}(2003)\citenamefont
  {Laustsen}, \citenamefont {Verstraete},\ and\ \citenamefont
  {Van~Enk}}]{Laustsen2003}%
  \BibitemOpen
  \bibfield  {author} {\bibinfo {author} {\bibfnamefont {T.}~\bibnamefont
  {Laustsen}}, \bibinfo {author} {\bibfnamefont {F.}~\bibnamefont
  {Verstraete}}, \ and\ \bibinfo {author} {\bibfnamefont {S.}~\bibnamefont
  {Van~Enk}},\ }\href@noop {} {\bibfield  {journal} {\bibinfo  {journal}
  {Quantum Information and Computation}\ }\textbf {\bibinfo {volume} {3}},\
  \bibinfo {pages} {64} (\bibinfo {year} {2003})}\BibitemShut {NoStop}%
\bibitem [{\citenamefont {Yu}\ and\ \citenamefont {Song}(2007)}]{Yu2007}%
  \BibitemOpen
  \bibfield  {author} {\bibinfo {author} {\bibfnamefont {C.-s.}\ \bibnamefont
  {Yu}}\ and\ \bibinfo {author} {\bibfnamefont {H.-s.}\ \bibnamefont {Song}},\
  }\href {\doibase 10.1103/PhysRevA.76.022324} {\bibfield  {journal} {\bibinfo
  {journal} {Phys. Rev. A}\ }\textbf {\bibinfo {volume} {76}},\ \bibinfo
  {pages} {022324} (\bibinfo {year} {2007})}\BibitemShut {NoStop}%
\bibitem [{\citenamefont {Ac\'{i}n}\ \emph {et~al.}(2000)\citenamefont
  {Ac\'{i}n}, \citenamefont {Andrianov}, \citenamefont {Costa}, \citenamefont
  {Jan\'e}, \citenamefont {Latorre},\ and\ \citenamefont {Tarrach}}]{Acin2000}%
  \BibitemOpen
  \bibfield  {author} {\bibinfo {author} {\bibfnamefont {A.}~\bibnamefont
  {Ac\'{i}n}}, \bibinfo {author} {\bibfnamefont {A.}~\bibnamefont {Andrianov}},
  \bibinfo {author} {\bibfnamefont {L.}~\bibnamefont {Costa}}, \bibinfo
  {author} {\bibfnamefont {E.}~\bibnamefont {Jan\'e}}, \bibinfo {author}
  {\bibfnamefont {J.~I.}\ \bibnamefont {Latorre}}, \ and\ \bibinfo {author}
  {\bibfnamefont {R.}~\bibnamefont {Tarrach}},\ }\href {\doibase
  10.1103/PhysRevLett.85.1560} {\bibfield  {journal} {\bibinfo  {journal}
  {Phys. Rev. Lett.}\ }\textbf {\bibinfo {volume} {85}},\ \bibinfo {pages}
  {1560} (\bibinfo {year} {2000})}\BibitemShut {NoStop}%
\bibitem [{\citenamefont {Tavakoli}\ \emph {et~al.}(2022)\citenamefont
  {Tavakoli}, \citenamefont {Pozas-Kerstjens}, \citenamefont {Luo},\ and\
  \citenamefont {Renou}}]{Tavakoli2022}%
  \BibitemOpen
  \bibfield  {author} {\bibinfo {author} {\bibfnamefont {A.}~\bibnamefont
  {Tavakoli}}, \bibinfo {author} {\bibfnamefont {A.}~\bibnamefont
  {Pozas-Kerstjens}}, \bibinfo {author} {\bibfnamefont {M.-X.}\ \bibnamefont
  {Luo}}, \ and\ \bibinfo {author} {\bibfnamefont {M.-O.}\ \bibnamefont
  {Renou}},\ }\href {\doibase 10.1088/1361-6633/ac41bb} {\bibfield  {journal}
  {\bibinfo  {journal} {Reports on Progress in Physics}\ }\textbf {\bibinfo
  {volume} {85}},\ \bibinfo {pages} {056001} (\bibinfo {year}
  {2022})}\BibitemShut {NoStop}%
\bibitem [{\citenamefont {Xiang}\ \emph {et~al.}(2022)\citenamefont {Xiang},
  \citenamefont {Cheng}, \citenamefont {Gong}, \citenamefont {Ficek},\ and\
  \citenamefont {He}}]{Xiang2022}%
  \BibitemOpen
  \bibfield  {author} {\bibinfo {author} {\bibfnamefont {Y.}~\bibnamefont
  {Xiang}}, \bibinfo {author} {\bibfnamefont {S.}~\bibnamefont {Cheng}},
  \bibinfo {author} {\bibfnamefont {Q.}~\bibnamefont {Gong}}, \bibinfo {author}
  {\bibfnamefont {Z.}~\bibnamefont {Ficek}}, \ and\ \bibinfo {author}
  {\bibfnamefont {Q.}~\bibnamefont {He}},\ }\href {\doibase
  10.1103/PRXQuantum.3.030102} {\bibfield  {journal} {\bibinfo  {journal} {PRX
  Quantum}\ }\textbf {\bibinfo {volume} {3}},\ \bibinfo {pages} {030102}
  (\bibinfo {year} {2022})}\BibitemShut {NoStop}%
\end{thebibliography}%

\end{document}